\documentclass[runningheads]{llncs}
\usepackage[T1]{fontenc}
\usepackage{graphicx}
\usepackage{latexsym}
\usepackage{tikz}
\usepackage[utf8]{inputenc}
\usepackage{xcolor}
\usepackage{amsfonts}
\usepackage{hyperref}
\usepackage{breakcites}

\usepackage{fancyhdr}

\begin{document}
\title{Understanding Privacy By Formalizing It\thanks{This work was supported by the Fonds National de la Recherche Luxembourg through the project Deontic Logic for Epistemic Rights (OPEN O20/14776480).}}
\author{Réka Markovich\inst{1}
\and
Truls Pedersen\inst{2}
\and
Marija Slavkovik\inst{2}}
%
\authorrunning{R. Markovich et al.}
%
\institute{University of Luxembourg
\email{reka.markovich@uni.lu}\\
\and
University of Bergen\\
\email{\{Marija.Slavkovik,Truls.Pedersen\}@uib.no}}
\maketitle              
\begin{abstract}
In most of the modern societies, there is a broad consensus regarding the need for promoting privacy and thus placing restrictions on technological---including AI---developments to protect people's right to privacy. In order to meet these expectations on the algorithmic level, first we need to make the concept of privacy and the related or derived rights formally specified. However, the notion of (the right to) privacy is subject to different interpretations. In this paper, we use a multi-modal logic to provide an initial formalization of different theories and approaches' basic principles and their implications investigating the right to privacy as an epistemic right within the theory of normative positions.
\end{abstract}

\section{Introduction}
In the context of ethical impact of artificial intelligence, privacy is often discussed as a value eroded by digitization and artificial intelligence \cite{ForbrukerradetNO}. Privacy, however, is not one of the traditional moral values \cite{Quine1978,Kinnier2000,Floridi2019Unified}. There are numerous attempts in the literature on defining privacy, but there is no consensus \cite{Matzner2019}. The overall privacy situation is made more confusing by what \cite{Elvy2017} calls ``emerging personal data economy''. The data economy both exploits and drives the need for more specific  privacy regulations.

On the global scale, privacy is a culturally divisive value or reference point, but in the so-called western countries there seems to be a broad consensus regarding its primary importance. This 
involves a vast regulative aspiration aiming at reasonable restrictions on the different technological developments. 
  We believe that on the long run, the implementation of regulative expectations or imposing self-regulative restrictions, will require formal specification on what privacy can mean and what the right to privacy exactly is. If we take seriously the numerous claims that artificial intelligence in particular, and digitization in general, undermine the existences of privacy, then we need to have a good understanding of what privacy is, what duties the right to it implies, and how it can be preserved. 

Algorithmic processes run faster and are more ubiquitous than human processing capabilities. If  privacy were a right to be guaranteed to users of digital technologies, we need to understand its specific scope, motivation and eligible trade-offs. If privacy were  a value  with which we need to align those algorithmic processes, we need to specify it mathematically  to the level that we can construct an algorithm that detects whether privacy is violated.  Our intended contribution is to use logic specifications both as a goal but also as a method to clarify the distinction between different concepts of privacy. Our aim in this project is to make privacy specifications accessible for algorithmic analysis. Only with the precision of logic specification we can compare two privacy conceptualisations and know whether they refer to the same or different ideas.

In this paper, we have aimed for formalization using a multi-modal logic in which we can accommodate the basic principles of \emph{some} of the different approaches, definitions of the right to privacy and then reason with them. Since we are interested in the different deontic consequences of each approach, for a logical-legal theoretical background, we use the theory of normative positions. Our aim is to show the variety with formal conceptual analysis, we do not provide meta-logical results. First we look into some privacy definitions, then shortly into the theory of normative positions (readers being familiar with the latter can skip that section). After those, first we introduce the language and then we provide formal representations of the different rights included or implied by the privacy approaches.

\section{Approaches to and Definitions of Privacy}

In this section we outline briefly the relation between the idea of privacy and how it is reflected different ``computational'' domains where privacy is discussed. We then discuss some definitions of privacy that have been influential in the past in law and social science and which we choose to focus on in our specification efforts later in this paper. As it will be obvious from the definitions we consider, we limit ourselves to privacy that a person can enjoy with respect to information about one self. In addition to informational privacy, one can discuss spatial privacy, bodily privacy, privacy of decisions etc.


 Privacy is colloquially equated with the concern of how personal data is handled\footnote{for a definition of personal data see \cite{GDPR}}. This is particularly the case in the context of data processing, including collection. Privacy is a long known and studied concept in cybersecurity. The field of {\em differential privacy} \cite{Dwork2008}, for example,  is concerned with methods for sharing data sets without making individuals identifiable in them.  The perception of privacy as concern for how personal data is handled sometimes also ``bleeds'' into the field of artificial intelligence, where also sometimes is equated to security issues regarding data access \cite{Liu2022}.  
 

 
  It is not particularly clear in the literature what impact AI directly has on privacy if we expand beyond the data security concerns.  AI-constructed behavior prediction helps identify patterns in private and, what we can call personal, data \cite{SlavkovikSPA21}. \cite{Ackerman2001} emphasize that ``privacy is maintained by allowing the user to disseminate only the necessary data, which cannot be used to identify the user''. It can be argued, however, that it is the data collection itself and the use of the analysis done by AI that directly contribute towards reducing the users' rights to control which information is available to whom and whose scrutiny is allowed. What machine learning does is find patterns in data. Data patterns allow us to infer information that is not explicitly available, and which might be information that someone is unwilling to share about themselves. AI, as it is today, does not erode privacy as part of its operation. It is \emph{how AI is used} rather than \emph{what it does}---that is the issue at hand. 

Given that we are interested in the normative space of actions that privacy implies, we investigate what the \emph{right to privacy} means. Definitions of what elements this right contains and what duties it entails vary and have developed over the years in different contexts \cite{Matzner2019}. Privacy can be seen as the right to be left alone, or to be exempt from unwanted scrutiny \cite{Rossler2005}---or, as often referred to in US case law, freedom from unwarranted publicity\footnote{For instance, Hogin v. Cottingham, 533 So. 2d 525 (Ala. 1988) citing Norris v. Moskin Stores, Inc., 272 Ala. 174, 132 So. 2d 321 (1961)}---or, for instance, to be exempt from social interaction \cite{Schwartz1968}. It is widely recognized that privacy is both beneficial for personal and social development \cite{Margulis2003,Rossler2005} and affected by the ease of information creation and processing enabled by digitization \cite{Schwartz2004}. There are efforts to help the denizens of the digital world to understand the implications of their own activities on their own privacy, however without a consensus on what information is relevant and how it should be communicated \cite{Barth2022}.

 Different authors have argued different privacy perspectives over time, and it can be argued that the right to privacy has evolved as we have evolved as a society. It is not our intention in this section to provide a systematic review of all the privacy definitions. We can and do only focus on few privacy definitions. 

We start with \cite{Warren1890} that provide one of the first and very influential legal definitions of privacy as the ``right to be let alone''. The definition of \cite{Warren1890} comes in the time when photography and printed press begin to impose on people's lives \cite{Matzner2019}. In the context of information, we can interpret it as the right that certain information about an individual is not accessible to anyone in any circumstances (contexts).

In 2023, one can argue, the modern human owner of a smartphone is never alone.  We can connect to other people instantaneously via the internet, but when we do that we leave digital traces that reveal very much about us \cite{Stachl2020}. 
Not being alone does not directly mean being without privacy. Already in 1968, \cite{Westin:1967} proposed that privacy can be seen ``as the claim of [individuals] to determine for themselves when, how, and to what extent information about them is communicated to others.''  This relaxes the privacy definition beyond the simple ``no access'' to the ``no access without permission'', namely  to the requirement that access of particular information (in all contexts) is in the hands of the person who is the subject of that information.  

\cite{Nissenbaum2009} argues that this idea of access with control is problematic. Information about one self can be freely permitted under some circumstances but not others. For example, while one can be happy to disclose one's HIV status in a dating app, the same permission cannot be taken to hold outside of that specific context\footnote{The dating app Grinder was fined 6.2 million euros in 2021 for such a violation  \hyperlink{https://www.forbrukerradet.no/side/grindr-hit-with-e-6-2-million-fine-in-response-to-complaint-from-the-norwegian-consumer-council/}{https://www.forbrukerradet.no/side/grindr-hit-with-e-6-2-million-fine-in-response-to-complaint-from-the-norwegian-consumer-council/}.}, for example to hiring agents and insurance companies even if the app is open to everyone. In addition to access and control, she would also specify context. The privacy requirement  becomes access with permission in a particular context for a particular aim.
 
On a somewhat orthogonal dimension \cite{Rossler2005} argues that what is problematic about access and sharing of information is based on what one is allowed to do with that information. Namely, the problem lies in using information about someone to stigmatise and scrutinise that individual who should have had the right to privacy. The motivation from this consideration comes from constraints that prevent someone to be alone in their private activities such as for example a disability or limitation of available resources (space, time, funds etc.). Therefore \cite{Rossler2005} argues that privacy can be understood as 
a ‘space’ where one can act without unwanted public scrutiny. The purpose of affording this freedom from public scrutiny is to preserve the autonomy and freedom of the individual. 


 





We are working with {\em private information} that we rather loosely---and only informally---define as information about an individual which that individual is not comfortable with being collected, processed, shared, known, used, accessed etc. by others. Private information overlaps, but may not necessarily subsume personal data as defined in the \cite{GDPR}. Private information in this sense is not only subjective but also contextual: the same information can be private in one context (for some people) but not in another. Within this work we abstract from the context details for the purpose of building up to capturing differences between the different strengths of epistemic privacy requirements.

Lastly we should clarify that influential taxonomies of privacy do exist, although somewhat dated, such as the one of \cite{Solove2006}. Solove bases his taxonomy on activities that invade privacy. Activities that invade privacy are arguably easier to discern by a human judge that wants to determine if privacy is violated.  However we are concerned with privacy erosion that occurs because of the contemporary capabilities of data science and AI. To this end we focus on the epistemic aspects of privacy and we take an epistemic approach to our specifications in logic. 

The \emph{first step} on the road leading to enabling specification is presented in this paper where we use a multi-modal logic for the \emph{formal conceptual characterization} of possible approaches to what the right to privacy means.

\section{Rights in the Theory of Normative Positions}
As a background framework, we use the theory which aims at formally differentiating in the different types of rights: the theory of normative positions \cite{Sergot}. Its origin is the paper of W.N. Hohfeld who found that lawyers overuse the word `right' meaning different concepts without even reflecting on it \cite{Hohfeld}. To resolve this terminological and thus conceptual confusion, Hohfeld proposed to differentiate the following four types of rights and their correlative duties (for details see \cite{MarkovichSL}):
\\
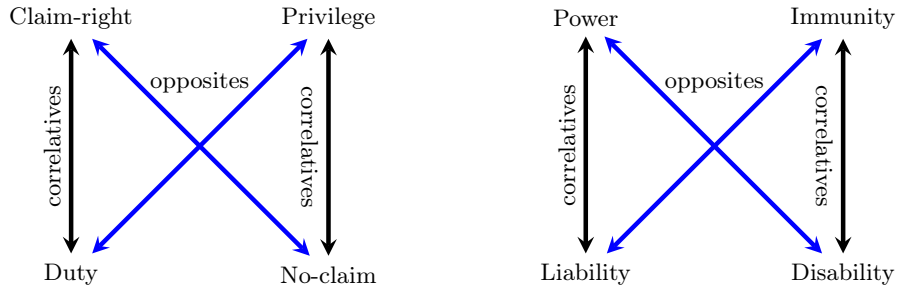
\begin{figure}[h]
    \centering
\begin{tikzpicture}[scale=3.4, >=stealth, line width=.6mm]
\begin{scope}[]
\node (v1) at (0,0) {Duty};
\node (v2) at (0,1) {Claim-right};
\node (v3) at (1,1) {Privilege};
\node (v4) at (1,0) {No-claim};
\draw[<->]  (v1) -- (v2) node[midway, above, sloped]{\footnotesize correlatives};
\draw[<->]  (v3) -- (v4) node[midway, below, sloped, rotate=180]{\footnotesize correlatives};
\draw[<->, blue]  (v2) -- (v4);
\draw[<->, blue]  (v1) -- (v3);
\node at (0.5,0.75) {\footnotesize opposites};
\end{scope}
\begin{scope}[xshift=2cm]
\node (v1) at (0,0) {Liability};
\node (v2) at (0,1) {Power};
\node (v3) at (1,1) {Immunity};
\node (v4) at (1,0) {Disability};
\draw[<->]  (v1) -- (v2) node[midway, above, sloped]{\footnotesize correlatives};
\draw[<->]  (v3) -- (v4) node[midway, below, sloped, rotate=180]{\footnotesize correlatives};
\draw[<->, blue]  (v2) -- (v4);
\draw[<->, blue]  (v1) -- (v3);
\node at (0.5,0.75) {\footnotesize opposites};
\end{scope}
\end{tikzpicture}    
    \caption{The Hohfeldian atomic types of rights, and their correlatives}
    \label{fig:hohfeldsquareds}
\end{figure}
A claim-right of an agent concerns the counter-party's actions. The counter-party has an obligation to do the certain thing, and this obligation is directed to the right-holder. Hohfeld calls this a duty, in the narrow sense. The freedom or privilege\footnote{`Freedom' is an often used alternative for `privilege' in the literature dealing with Hohfeld.} to do something, on the other hand, is understood as not being the subject of a claim-right coming from the counter-party. Privilege can thus be seen as a directed version of the standard (weak) permission in deontic logic.

The normative positions in the right square capture the agent’s ability to change an (other) agent's normative positions. For that reason, they have been called higher-order or \textit{capacitative}~\cite{Fitch}. They thus capture the norm-changing potential---or in case of disability, the lack thereof---of an agent~\cite{OlivierHuiminJournalversion,MarkovichSL}. The counterparty's exposedness to this change is called liability in this theory, while immunity is the type of right when there is no such exposedness since the other agent has no power.

The theory of normative positions covers the tradition of aiming at formalizing these positions established by the work of Kanger and Kanger (e.g.\cite{KangerandKanger,Kanger1972}) and Lindahl (e.g.\cite{Lindahl1977}), and later joined by many (e.g. \cite{Makinson,Sergot,Sartor,MarkovichSL,OlivierHuiminJournalversion})

\subsection{Rights as Absolute Positions}
One of the main characteristics of the theory of Hohfeld is that it considers agents pairwise, as \cite{Makinson} put it: it is inherently relational. This aspect is often source of criticisms allegedly lead a narrow scope: considering agents pairwise is a good tool to describe contractual situations, but inadequate for the so-called absolute positions (like, for instance, the ownership). This critique is ill-founded. Hohfeld's famous essay on the fundamental legal conceptions had a second part in which he differentiates between the \emph{paucital} and \emph{multital} rights in the case each right-type. A paucital right-relation refers to situations in which we indeed have one-one agent on each side of the relation, like in the case of contracts. Multital rights are, however, series of such relations where one agent takes the right-holder positions and everyone else is a duty-bearer regarding her rights. As the example of \cite{Simmonds} shows: ``I am the owner of Blackacre. I have a claim-right that you should not enter the land without my consent. I have the identical claim-right against your mother, my employer, the Bishop of Ely, and anyone else that you care to mention. Each of these
claim-rights is a consequence of my ownership of Blackacre. These are ‘multital rights’.'' The same is true for the owner's freedom to walk on his own land: it means that no one has a claim against him to refrain from walking through it, which is a multital freedom; the owner has a multital power to sell this land: everyone is exposed to the change this sale bring in their normative positions; and the owner has a multital immunity too as everyone else is disable to sell his land changing his normative positions about it. 

This addition has great relevance in using this theory for modeling privacy rights: many of the rights privacy implies seems to refer to a unique position that we have against everyone else. We will represent it as a conjunction of relations between one agent and each of the others on the (given) set of agents.

\subsection{Rights as Molecular Normative Positions}
Hohfeld's initiative didn't succeed in the sense that people---lawyers too---still use the word 'right' without special reflection to what exactly they mean. From the analysis of the attached regulation one can mostly figure out which normative position is covered by the expression. For instance, as pointed out in \cite{markovichformalizing}, in Hungary the citizens have a right to know the declaration of assets both in the case of MPs and the local representatives. But these two rights are different: the actual regulation orders the declarations of the MPs to be submitted and made public, while in the case of the local representatives, the law says that in case of a citizen's inquiry, the representative is obliged to submit her declaration to be made public. That is, the first right is a claim-right, the second is a power. However, rights are often refer to not even one atomic position, that is, one of the above mentioned four right positions, but a molecular one. The broader the context, the more probable case is that the right we talk about is a complex one. In case of the human rights, establishing some fundamental interests and their protection, this is rather probable. See, for instance, the logical analysis of freedom of thought \cite{FoT}. We believe that the right to privacy is often understood as a combination of different atomic rights positions.

\subsection{Epistemic Rights}
The notion of epistemic rights as a robust category and investigation of them within the Hohfeldian theory was established in epistemology by \cite{Watsonbook}. Watson condsiders epistemic rights as those protecting and governing the distribution and accessibility of epistemic goods. Developing (and extending existing) logics for formalizing epistemic rights within the theory of normative positions was initiated by \cite{FoT} and \cite{RekaOlivier2021}. In these papers we find a differentiation between epistemic rights in the narrow and the broad sense referring to the theory of normative positions. According to this, epistemic rights in the narrow sense are those that concern the right-holder's epistemic state, like the right to know or the freedom of belief. In the broad sense though, those rights are also epistemic rights that concern the duty-bearer's epistemic state, like the right to be forgotten and the different rights to privacy. Hence, in this paper, we investigate the formalization of these latter as epistemic rights within the theory of normative positions. 

\section{Language and Semantics}
For the formal characterization of these epistemic rights, we use a combination of standard deontic logic augmented with directed operators~\cite{MarkovichSL}, action, epistemic, and alethic logic. We are going to work with the following multi-modal language:

\begin{definition}
Let $A$ be a finite set of agents and $\Phi$ a set of propositional letters. The language $\mathcal{L}$ is defined as follows:
\[
p\in \Phi\mid \phi\land\psi \mid \neg\phi \mid \Box\phi \mid \{ E_a\phi \mid O_{a\to b}\phi \mid K_a\phi \}_{a, b \in A} \]
\end{definition}
$\mathcal{L}$ thus extends the propositional logic with four modalities. $E_a$ is the agency modality and should be read as ``agent $a$ sees to it that...''. $O_{a \to b}$ is a directed obligation modality, and should be read as ``agent $a$ has a duty towards $b$ that...''. $K_a$, on the other hand, is an epistemic modality, to be read as ``agent $a$ knows that...''. The $\Box$ is the alethic modality ``it is necessary that''.  All these modalities have duals: the weak permissions operator, i.e. $P_{a \to b}...$, which stands for $\neg O_{a\to b} \neg...$; $\langle K_a \rangle ... $ which stands for $\neg K_a \neg ... $; and $\Diamond ...$, which stands for $\neg \Box \neg...$.

We make the following assumptions regarding the logical behavior of these modalities. We take the deontic modalities $O_{a \to b}$ to be normal modalities validating the D axiom, i.e. $O_{a \to b} \phi \rightarrow P_{a \to b} \phi$. So the deontic fragment of our language is standard deontic logic. In this paper we work with a very weak action logic, so we take the agentive modalities $E_a$ to be non-normal, validating only substitution under logical equivalence and the $T$ axiom ($E_a \phi \rightarrow \phi$). 
The epistemic modality $K_a$ is assumed to be normal modality validating the T,4,5 (and thus the B) axioms, that is, we choose $K_a$ to be the standard knowledge modality. The $\Box$ operator refers to a modality satisfying also K,T,4, and 5. Given these assumptions, the language $\mathcal{L}$ is interpreted over frames containing a neighborhood function for each $E_a$, a deontic ideality relation for each $O_{a \to b}$, an epistemic accessibility relation for each $K_a$ and an alethic accessibility relation. 
\begin{definition}
Let $A$ be a finite set of agents. A frame $\mathcal{F}$ is a tuple of the following form:
\[  
\mathcal{F} = \langle W, R^{\Box}, \{f_{a}, R^{K}_a, R^{O}_{a,b} \}_{a,b\in A} \rangle
\]
Here $W$ is set of possible worlds. The function $f_a:W \to \wp\wp(W)$ is a neighborhood function such that, for all $w \in W$ and  $X \in f_a(w)$, we have $w \in X$. $R^{O}_{a,b} \subseteq W^2$ is a serial binary relation. $R^{K}_a \subseteq W^2$ and $R^{\Box}\subseteq W^2$ are Euclidean relations. A model $\mathcal{M}$ is a frame $\mathcal{F}$ together with a valuation function $ V: \Phi\to\wp(W)$.
\end{definition}

\noindent With this in hand the truth conditions of formula of our language is defined in the standard way. We have only defined explicitly the case for the modalities.

\begin{definition}
Let  $||\phi|| = \{w : \mathcal{M}, w \models \phi\}$. Then:
\begin{itemize}
\item $\mathcal{M}, w \models \Box\varphi \Leftrightarrow \forall w'(wR^{\Box} w'\Rightarrow \mathcal{M}, w' \models\varphi)$
\item $\mathcal{M}, w \models E_a\varphi \Leftrightarrow ||\phi|| \in f_a(w)$
\item $\mathcal{M}, w \models \mathbf{O}_{a\to b}\varphi \Leftrightarrow \forall w' (w R^{O}_{a,b} w' \Rightarrow  \mathcal{M}, w'\models \varphi)$
\item $\mathcal{M}, w \models K_a \varphi \Leftrightarrow \forall w' (wR^{K}_a w'\Rightarrow \mathcal{M},v \models\varphi)$
\end{itemize}
\end{definition}
 These truth conditions are standard for the normal modalities $K_a$, $O_{a\rightarrow b}$, and $\Box$. The agency operator $E_a$ is given the so-called \emph{exact} neighborhood semantics~\cite{pacuit2017neighborhood}. Validity in models, frames, and classes thereof, are defined as usual. Since we do not make any specific assumptions regarding the interaction between these modalities, the set of validities over our intended class of frames is completely axiomatized by all propositional tautologies, the logic ET for the agentive modality $E_a$, KD for $O_{a \to b}$, and S5 both for $K_a$ and $\Box$.

\subsection{Motivation of the Language}
We chose to use this language to be able to express different variants of what the right to privacy might mean. The directed obligation refers to the Hohfeldian duty emphasizing the relationality, which will always have an $E$ operator in its scope (however, for a staring point, we show below formulas with an undirected obligation too). The very weak action logic enables to talk about ``actions'' in a very broad sense and even iterate the operator (which would not be so easy with a usual S4 or S5 STIT logics) without engaging with the substantial questions of what actually actions are. We chose S5 though to `knowledge'. We are aware that the adequate choice of axioms for properly characterizing knowledge is extensively discussed, and we do not intend to take position with our choice. At this phase of the current research project we put the emphasis of the finding the formulas expressing the variant of the rights related to privacy\footnote{In a later phase of this research, we will modify the logic according the findings, such as counterintuitive consequences in a given context, using different modalities as the epistemic notions involved in the discussion about privacy might greatly vary.} Using the different combinations we intend to express some basic components of (privacy-related) actions and positions, such as $\Diamond E_a \phi$ as an ability to make it the case that $\phi$, $\Diamond E_a O_{b\rightarrow a} E_b \phi$ as having the power to put a duty on $b$ to make it the case that $\phi$. The formula $\Diamond K_a\phi$ is intended as $a$ has access to $\phi$, $E_b\Diamond K_a \phi$ as $b$ making $\phi$ accessible for $a$ as opposed to $E_b K_a \phi$ as $b$ making $a$ know $\phi$. The modularity of the combinations enables us to express seemingly only slightly different concepts which however might have very different consequences. To have a simple language and since we always operate with a finite set of agents, we choose to stay in propositional modal logic.

\section{Formalization of Some Right to Privacy Definitions}

\subsection{Right to be left alone: the right to control who has access}
To say that agent $a$ has to right to make it the case that others ($b$ such that $b\neq a$) do not know some information ($\phi$)---as in it should be the case---seems to be a legit starting point:
\begin{equation}
    O\Diamond E_a\neg K_b\phi
\end{equation}
It is somewhat different to say is that agent $a$ has to right to make it the case that others cannot know (do not have access to) some information:
\begin{equation}
    O\Diamond E_a\neg\Diamond K_b\phi
\end{equation}
The two formulas above are `ought-to-be' formulas, they do not express rights directly. In order to fit the theory of normative positions the agents of the normative relations have to be specified. Formally this can be done with the obligation operator being indexed with a pair of agents as introduced in \cite{MarkovichSL}. In order to have `ought-to-do' formulas, an action operator have to be in the argument of the obligation operator indexed with the obligation's first indexed agent. The obvious candidate for creating such a situation is the state (legislator). It seems to be plausible to say for some specified set of formulas, it is the state's duty to make it the case that an agent can decide about the publicity of $\phi$. Actually, if we accept that it is a state duty, then it is regarding each of its citizen:
\begin{equation}
    \bigwedge\limits_{a,b\in A}O_{s\rightarrow a}E_s \Diamond E_a \neg \Diamond K_b\phi
\end{equation}
Actually, this requirement might be too strong toward the state. The legislator's tool is rulemaking, not implementing technical constraint (not to mention metaphysical ones). Thus it seems to be more appropriate to say the following:
\begin{equation}
    \bigwedge\limits_{a,b\in A}O_{s\rightarrow a}E_s O\Diamond E_a \neg \Diamond K_b\phi
\end{equation}
Such a legislation does not solve the problem yet as it does not identify the agent which has to make it the case. We need to point out a duty-bearer:
\begin{equation}
    \bigwedge\limits_{a,b\in A}O_{s\rightarrow a}E_s O_{c\rightarrow a} E_c \Diamond E_a \neg \Diamond K_b\phi
\end{equation}
The agent $c\in A$ can be a company, or any other agent that the state can impose such a duty on, where we also allow for $c=b$ (but we require $a \neq b$ and $a\neq c$). In the Hohfeldian terms, the formula above is a \emph{claim-right} of (every) $a$ against to state to establish a clam-right against the relevant company making it possible that $a$ can decide who knows $\phi$. According to the interpretation of \cite{Westin:1967} privacy is attained by a person when that person can control who can share and use their information thus including another related claim-right of $a$ realized by a duty of everyone else to refrain from making $a$ unable to let others know:
\begin{equation}
    \bigwedge\limits_{b,c\in A}O_{c\rightarrow a}\neg E_c \neg\Diamond\neg E_aK_b\phi
\end{equation}

The right to privacy definitely includes $a$'s  \emph{multital freedom} as well: that she does not have an obligation letting others know about $\phi$ (or that she even makes $\phi$ accessible to others):
\begin{equation}
    \bigwedge\limits_{b\in A}\neg O_{a\rightarrow b}E_a K_b \phi
\end{equation}
\begin{equation}
    \bigwedge\limits_{b\in A}\neg O_{a\rightarrow b}E_a \Diamond K_b \phi
\end{equation}
However, a freedom this way, in itself, is just a weak permission. It has to come with some protection to relaize what we usually mean by what a freedom is. The classical protection is what the Hohfeldian \emph{immunity} covers: the disability of other to change this freedom, which looks like the following in our formalization:
\begin{equation}
    \bigwedge\limits_{b\in A}\neg\Diamond E_b O_{a\rightarrow b}E_a \Diamond K_b \phi
\end{equation}

In the above cases we use the tacit assumption that it is possible that someone can be left alone in the metaphorical sense of having total control on the access to $\phi$. However, this is not always the case.

As \cite{Rossler2005} argued, being alone or access to control to private information may be unattainable in some circumstances or for some people. In such a case, we have to calculate with agents who do have access to $\phi$, and the rights to privacy are realized in some control in the normative space of these agents regarding $\phi$-related actions. So in cases where it is inevitable that $b$ has access to $\phi$, one obvious candidate for $a$'s privacy rights is that $a$ can prohibit (or permit) making $\phi$ :
\begin{equation}
    \bigwedge\limits_{b,c\in A}\Box((\Box\Diamond E_b K_b\phi)\rightarrow (\Diamond E_a O_{b\rightarrow a}\neg E_b \Diamond K_c\phi))
\end{equation}
It can be questioned whether in these situations it is indeed $b$ who sees to it that he knows $\phi$, thus the formula below might be found more accurate:
\begin{equation}
    \bigwedge\limits_{b,c\in A}\Box((\Box\Diamond K_b\phi)\rightarrow (\Diamond E_a O_{b\rightarrow a}\neg E_b \Diamond K_c\phi))
\end{equation}
It is an interesting interpretation of privacy as space without uninvited scrutiny to say that agent $a$ might not only want that further agents have no access to $\phi$ but maybe she has to be able to control whether she has to face uninvited opinions of those who necessarily have access to $\phi$. That is, even if $b$ as a helper or servant is necessarily witnesses $\phi$, $a$'s right to privacy covers that she prohibits that $b$ lets her know about this:
\begin{equation}
    \Box((\Box\Diamond K_b\phi)\rightarrow (\Diamond E_a O_{b\rightarrow a}\neg E_b K_a\phi))
\end{equation}
Note that this formula differs from the one above only in missing a $\Diamond$ and talking about $b$ letting know $a$ and not a third party.

\subsection{Right to transparency}
In some systems, the question is not really the total exclusion of any type of access (as it might not be feasible under all circumstances), but rather the right to transparency: agent $a$ should know about whether anyone has access to $\phi$. To express such a claim-right, that is, the directed obligation of the agent controlling the system, we apply the solution of \cite{Hulstijn08} of `knowing whether' which avoids the infamous \r{A}qvist's paradox~\cite{Aqvist} making it possible at the same time that we do not rely on conditionals:
\begin{equation}
    \bigwedge\limits_{b\in A}O_{c\rightarrow a}(E_c K_a \Diamond K_b \phi\lor E_c K_a \neg\Diamond K_b \phi)
\end{equation}

\subsection{Protection: possibility of enforcement}
An important aspect of our claim-rights, that is, the duties of others regarding our privacy is that once they are violated, we (on the metaphysical level) have a new claim-right against the judiciary to enforce or rights (or compensation for the violation) as described in detail and formalized in \cite{MarkovichSL}.
For instance, in case of the company's duty to enable the user to make $\phi$ inaccessible:
\begin{equation}
    \Box((\neg E_c \Diamond E_a \neg \Diamond K_b\phi)\rightarrow O_{j\rightarrow a} E_j E_c\Diamond E_a \neg \Diamond K_b \phi)
\end{equation}

This (on the practical level) means that we have a \emph{power} to initiate a legal action putting a duty on the judiciary to decide whether indeed that was the case what we state. This instrumental aspect is discussed in detail and formalized in \cite{MarkovichRoyJURIX21}, we do not go into details here.



\section{Discussion, Related and Future Work}
We have introduced a multi-modal logic to formalize some approaches of what the right to privacy means pointing out several different normative positions. This work brings together two aspects that have been present in computer science. On the one hand, the need for expressing privacy-related concepts have been addressed in the literature using logic for (legal) knowledge representation. \cite{ABT1} and \cite{ABT2}, in order to deal with privacy policies, investigated both the obligation and the permission to know, differentiating between obligatory and permitted knowledge and obligatory and permitted messages. \cite{Xu2022} use dynamic logic to describe permitted announcements. On the other hand, within the context of multi-agent systems, privacy is studied from several aspects such as: artificial  agents assisting people in maintaining their privacy \cite{ijcai-22-privacy}, identifying ``leaks'' of particular information \cite{DennisSF16}, negotiation to resolve privacy conflicts among people \cite{Such2016, toit-17-prinego}, preservation of privacy during learning \cite{Nagar2021}, to name a few. In these approaches, privacy is seen as different  property of states and/or actions, but not as an epistemic right. Both in logic and MAS, the works by different authors build on different understandings of privacy---our work aims exactly at making them comparable. Furthermore, since these considerations of privacy in logic and MAS are typically not grounded in the law and social sciences literature, it is also difficult to ground the them into the state-of-the-art outside of computer science.

Our work aims at providing foundations for a research going for implementable specifications of privacy-related rights. This paper provides and initial formal conceptual analysis contributing to legal knowledge representation, and to set a basis in which to ground privacy work AI, MAS, including policy modeling, policy-as-code  and law-as-code paradigms and initiatives. Among our next steps there are addressing the defeasibility of these rights, trying other formalisms, and using the LogiKEy  framework for the design and engineering of ethical reasoners, normative theories and deontic logics put forth by \cite{Logikey} to see which works best.

\bibliographystyle{apalike}
\bibliography{privacy_jurisin_2023}
\end{document}